\documentclass{article}

\usepackage{fancyhdr}
\usepackage{lipsum}

\usepackage{hyperref}

\hypersetup{
    colorlinks=true,
    linkcolor=blue, 
    urlcolor=blue,
    citecolor=blue
}

\usepackage{float}
\usepackage{multirow}
\usepackage{cite} 
\usepackage{graphicx} 
\usepackage{amsmath} 
\usepackage{tikz} 
\usepackage{pgfplots} 
\usepackage{booktabs} 

\usepackage[T1]{fontenc}
\usepackage{mathptmx}

\usepackage [autostyle, english = american]{csquotes}
\MakeOuterQuote{"}

\usetikzlibrary{patterns, intersections} 
\usepgfplotslibrary{fillbetween} 

\newcounter{myindex}

\newcommand{\Hindex}[1]{\refstepcounter{myindex}(H\arabic{myindex})\label{#1}}

\title{General collections demography model with multiple risks}
\author{Josep Grau-Bové and Miriam Andrews}
\date{June 2024}

\begin{document}

\maketitle

\begin{abstract}
This note presents an Agent-Based Model (ABM) with Monte Carlo sampling, designed to simulate the behaviour of a population of objects over time. The model incorporates damage functions with the risk parameters of the ABC framework to simulate adverse events. As a result, it combines continuous and probabilistic degradation. This hybrid approach allows us to study the emergent behavior of the system and explore the range of possible lifetimes of a collection. The main outcome of the model is the decay in condition of a collection as a consequence of all the combined degradation processes. The model is based on six hypotheses that are described for further testing. This paper presents a first attempt at an universal implementation of Collections Demography principles, with the hope that it will generate discussion and the identification of research gaps.
\end{abstract}

\section{Introduction}


The concept of Collections Demography developed by Strlic \cite{strlivc2015damage} originated from the need to develop evidence-based management strategies for historic collections. In the context of libraries and archives, users are primarily concerned with the loss of textual information, which can lead to degradation being deemed unacceptable and an object becoming unfit for use. To address this, researchers have developed damage functions that combine aspects of material degradation, use, and material attributes important for user interaction with heritage \cite{strlivc2013damage}. These functions are based on data from paper degradation experiments and real collections. Michalski also proposed a collection-wide model focused on mechanical damage to paintings \cite{michalski2013stuffing}. Collections demography has also been used for the evaluation of scenarios for managing storage environments and levels of access for different types of library and archival paper \cite{duran2021comparison}. Clearly, the collections demography principles have demonstrated advantages in the sustainable management of collections. However, their application has so far been restricted to very specific collection types. The question remains as to whether collections demography could be extended to any collection, from coins to churches?

Recent efforts by the Department for Culture, Media and Sport (DCMS) of the UK Government to develop a framework for valuing culture and heritage capital have highlighted the significance of damage functions in this context \cite{clark2021policy}. The goal of the "Culture and Heritage Capital Framework" is to create a formal approach to value culture and heritage assets, which will ultimately inform decision-making processes in the public and private sectors. Sagger and Bezzano have put forward a proposal to integrate economic valuation methodologies with degradation rates \cite{Sagger2024} to measure the welfare impact of interventions that halt the loss or deterioration of cultural and heritage assets. This requires an ability to calculate how any intervention, to any collection, will impact its lifetime.

The success of these policy initiatives is predicated on the existence of sufficiently accurate damage functions to predict the degradation of most relevant collections and buildings. This is complicated for two main reasons. First, there are only complete damage functions for a handful of materials. So far, successful applications that are able to accurately predict the lifetime of real collections (rather than only establishing degradation rates) are limited to the hydrolysis of cellulose. Secondly, damage functions that predict chemical degradation are only half the picture. As it is well known, heritage degrades through both continuous deterioration and catastrophic events \cite{michalski1990overall} . To accomplish the the ambitious vision of the "Culture and Heritage Capital Framework", the field of heritage science must come together to develop more comprehensive damage functions and to establish better understanding about the level of risk that cultural heritage institutions face, while  ensuring these two interrelated processes work together effectively. However, this ideal may be years away.

This paper proposes that, in the meantime, agent-based models or other similar statistical models hold the key to combining the different types of degradation processes that affect collections, as well as handling the uncertain behaviour of the system. In other fields, such methods have been widely used to study the ageing of different types of "populations",  from the literal ageing of patients  \cite{spijker2022impact}, to survival rates during clinical trials \cite{an2001agent}, to the mechanical breakdown of engineering assets as diverse as pipeline infrastructure \cite{li2020new} , structural components within civil engineering  \cite{guo2020two}, and maritime vessels \cite{liu2018time}. All these examples are partial analogies to heritage collections. They display some of the key features of similar to the deterioration system of cultural heritage, but rarely all of them. The characteristics that define this system can be considered as follows:

\begin{enumerate}
    \item There is a finite population
    \item Each agent in the population has a key property (such as condition or value) that decreases over time
    \item The decreasing property is affected by gradual rate-processes as well as probabilistic accidents
    \item The process of decay is extremely slow, of the order of centuries
    \item The slowness of the process is such that it cannot be assumed that the decreasing property (condition or value) will be defined in the same way in the future. In other words, societal change is of a similar time-scale than the physical process. 
\end{enumerate}

As far as the authors are aware, there are not any analogous or similar systems in any other field that fulfill all these criteria, and certainly none that have already been modelled using computational simulation. Hence, the application of ABMs to heritage collections provides a new modelling scenario and will, inevitably, be full of research challenges. This paper presents a first attempt at the problem, with the hope that it will generate discussion and the identification of research gaps. To aid discussion, every time we introduce a new hypothesis, it will be noted and marked with an index like "H0".

\section{An operational definition of lifetimes}

To calculate the lifetimes of heritage objects, a definition of unacceptable degradation is required. A limit that is commonly adopted is the threshold beyond which the social function of an artifact changes fundamentally, i.e., the 'end-of-life' \cite{strlivc2013damage}. For example, a book may be degraded to the point it cannot be handled by readers, or a tapestry may be faded to the point of not being decipherable. Another, alternative, definition can be understood as the point where value loss is such that a clear need for investment or action emerges. The word "operational" signifies clearly that this threshold is meant to enable informed management. 

A reductive approach is necessary in deciding a threshold because it help to inform better decision-making. However, it is adopted with the full knowledge that any quantitative straight line drawn over a social continuum is, of course, a fiction. In many instances, it is difficult to approximate this limit of unacceptable degradation because many stakeholders will perceive the damage differently \cite{taylor1999investigation}. What is damage in some contexts is patina in others. Social and curatorial contexts influence the perception of age and loss \cite{grossi2004aesthetics}. The fundamental hypothesis of collection demography is that these thresholds are definable for a wide diversity of heritage typologies \Hindex{index:exist}. Only one or two decades of research in perception and social value of heritage may bring us closer to workable definitions of unacceptable change for all the useful cases.

\begin{figure}[h]
    \centering
    \includegraphics[width=1\textwidth]{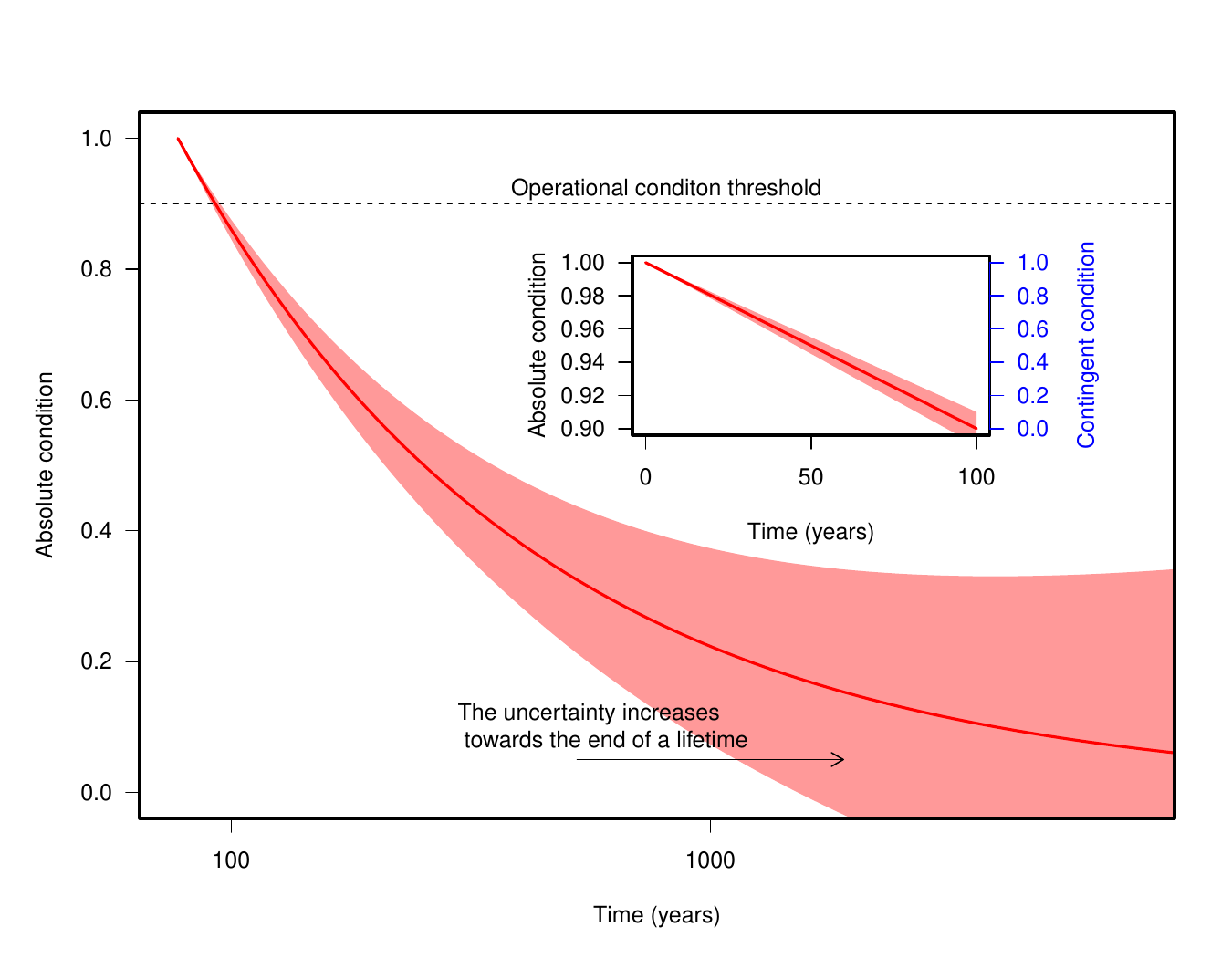}
    \caption{Condition decays in an unknown and unpredictable way. But the crux of decision making is focused on a small time window at the start of the process.}
    \label{fig:explanatory}
\end{figure}

These issues may be set aside in the pursuit of a practical solution. Within this scenario, it is necessary to assume that a condition state can be defined for each artifact: a quantity that decreases from 1 to 0 as the artifact ages. This is purely a theoretical construct and can be referred to as the absolute condition of the object. In practice, there is no need to attempt to model the entire lifetime of an object, from condition 1 to condition 0, for two reasons. Firstly, it is likely than in a lifetime of hundreds of years, the way society sees an object will change more than the object itself. Secondly, it is likely that as an object approaches the end of its lifetime, its function will change, for example through deaccessioning, joining a handling collection, or being forgotten. When this happens, the artifact will drop out of the management system we are attempting to model. To avoid the great uncertainty that these processes will bring to modelling, it is sensible to refrain from predicting the decay of absolute condition. Let us consider instead that, at some point early in the lifetime, the absolute condition will hit the operational limit that will trigger an action or investment. For example, as commonly assumed, a museum will worry when fading begins to be visible, rather than when a watercolor is absolutely white. As Figure \ref{fig:explanatory} represents, this point in time helps us define a shorter, less uncertain lifetime. We define a new contingent condition as the one that becomes 0 when the absolute condition hits the operational threshold. 

We should note that in this simple framework we are introducing another important dynamic hypothesis: that the probability of an action in response to condition increases in a step-wise manner when condition decreases. We can set up a threshold of unacceptable condition with ease when there is a condition above which most actors perceive a change or decide to act upon it. The sharper this change in perception, the more realistic it is to define an operational threshold. While a step-wise response is not essential, it is highly desirable. We may call this the "sharp response hypothesis" \Hindex{index:first}. In some cases, for example, the number of observers who see a surface as visibly dirty has been found to increase sharply at 5-10$\%$ area coverage by particulates. However, this behaviour may not be universal.

\section{The model}

\subsection{Modelling degradation rates with damage functions}

\begin{figure}[h]
    \centering
    \includegraphics[width=0.9\textwidth]{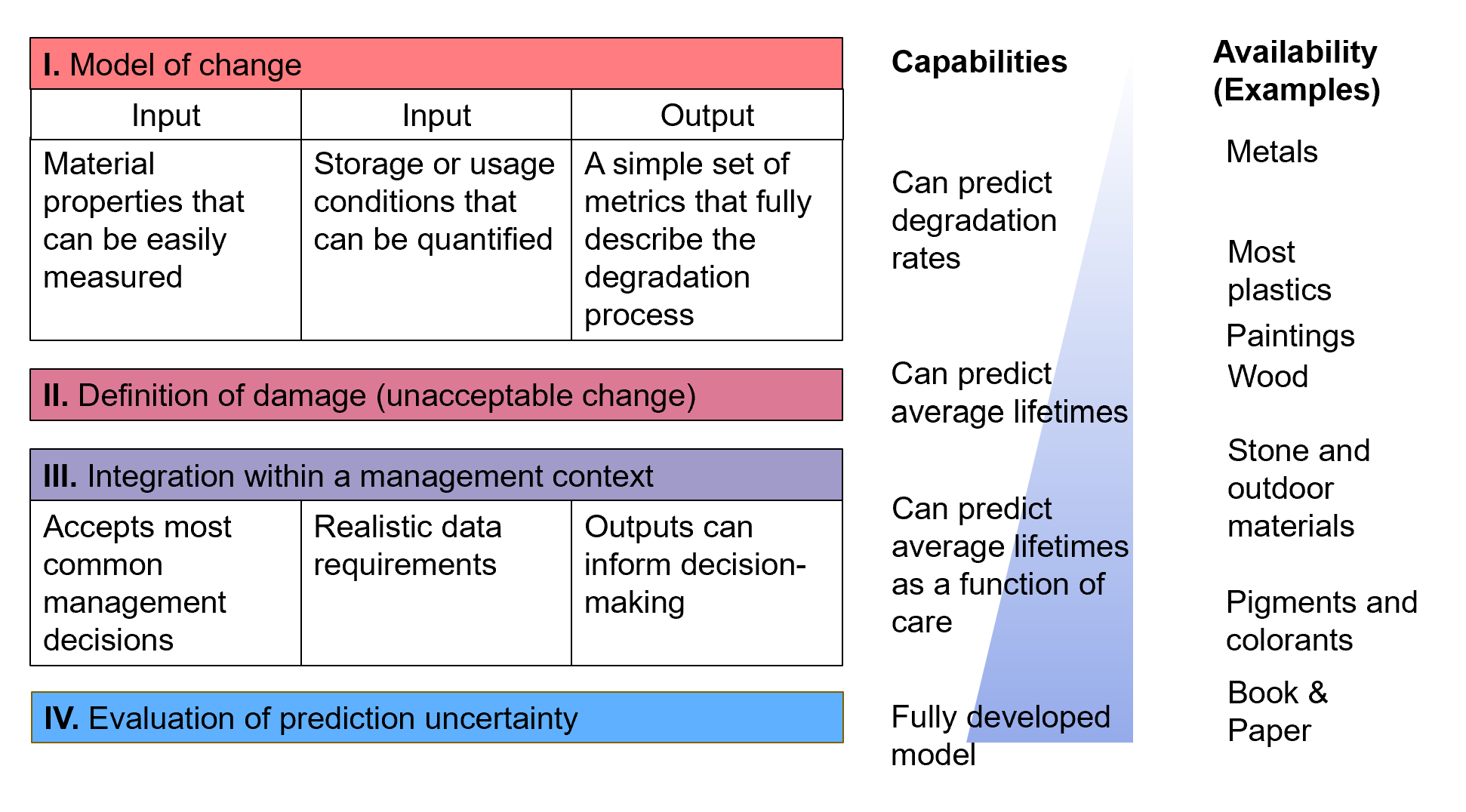}
    \caption{Summary of the capabilities of damage functions depending on their level of development.}
    \label{fig:summary}
\end{figure}

The general demographics model relies on the existence of damage functions that provide the decay rate of key properties (color, surface recession, mechanical strength, degree of polymerisation, etc.). It is necessary to  assume that these key properties are directly proportional to the absolute condition of an object \Hindex{index:proportional}. The model also requires a definition of the operational threshold for these properties as an input. However, not all damage functions are at the required level of development. For the purpose of clarifying the state of development of damage functions for different materials, we gave classifiedthem into four levels below (Figure \ref{fig:summary}):

A Level I damage function includes a model of change with well-defined inputs and outputs, identifying which input parameters are critical and which can be ignored. There are models at this level of detail for many materials, for example, for metals, cellulose acetate, or paintings on canvas and wood. At this level, we understand the physicochemical processes at play.

A Level II damage function can estimate lifetimes, because the physicochemical model is paired with a definition of damage derived from stakeholder preferences and value judgments. For example, we know at which point fading is just noticeable. It is also possible to define levels of stone recession where essential detail is lost. Many heritage materials such as plastics, paintings, and wooden objects are nearing level II, although the definitions of damage for most remain open to debate. 

Level III damage functions have aligned their input with management decisions. They contain inputs that correspond with what a manager would know. The functions that predict fading are good examples, as they relate to lux and light spectra of common sources.

The most advanced Level IV damage functions also evaluate uncertainty, offering lifetime estimates with a margin of error. Currently, only the damage functions for paper have reached this level. 

The model presented here requires at least Level II damage functions.

\subsection{Definition of adverse events}

This model takes advantage of the ABC framework for risk assessment \cite{iccrom_abc_method}. Consequently, adverse events in the model are defined by three parameters, each characterized by a level of uncertainty.

\begin{itemize}
    \item \textbf{Fraction of Collection Affected:} This parameter represents the proportion of the population or collection of objects that is impacted by an adverse event. It is expressed as a range, such as 0.01 to 0.2, indicating the uncertainty in the extent of the event's impact.
    
    \item \textbf{Extent of Impact on Condition of Objects:} This parameter quantifies the degree to which the condition of affected objects is reduced following an adverse event. It is given as a range, such as 20 to 40.
    
    \item \textbf{Mean Time:} This parameter defines the mean time before an adverse event occurs. It is also expressed as a range, reflecting the uncertainty in the timing of event occurrences.
    
\end{itemize}

The application of the ABC risk assessment model in this context approaches the limit of its suitability and intended use. This ABC model is primarily designed to facilitate decision-making by comparatively identifying and prioritizing the most significant risks in a semi-quantitative scoring system, rather than delivering precise risk predictions. The authors of this type of risk assessment have cautioned users about its limits from the very start. The words of Robert Waller in 1994 are still true: "Currently, the information required to produce accurate estimates of the magnitude of many risks is lacking. Nevertheless, simply attempting the exercise among a group of collections care staff produces several valuable results"  \cite{waller1994conservation}. Another critical hypothesis is, therefore, that the ABC method can eventually become the basis of quantitative forecasts \Hindex{index:ABC}. As we shall see, the simulation model presented in this paper can help evaluate this hypothesis. 

Table \ref{tab:degradation_processes} contains several examples of some of the types of adverse events that pose a threat to cultural heritage. Due to the catastrophic nature of fire events and the heightened number of incidences within historic buildings \cite{Landis2017}, fire often features within many risk and vulnerability assessments for cultural heritage \cite{ashley2013risk} \cite{salazar2021review} \cite{ulucc2022fire}. The availability of statistics regarding the frequency of fires is relatively good in some places \cite{kincaid2022fire}, with potential incidence rates for different levels of prevention, mitigation, and control measures also beginning to be defined \cite{tetreault2008fire}. 

Like fire, the theft of objects can also constitute another form of total value loss to an artifact since, in the words of the National Trust Manual of Housekeeping, "a stolen object is, to all intents and purposes, as lost as one that is destroyed in a fire" \cite{NationalTrust2011}. The high incidence of heritage crime \cite{HECrime} means that security threats are another focus of risk assessments for cultural heritage \cite{NERisk}. However, more sector-specific statistics need to be developed to understand the risk levels for museums with different combinations of crime prevention measures. 

Flooding is a common concern for heritage managers due to the physical damage and staining caused by exposure to water, as well as the mould infestations that often emerge upon previously submerged building components, and has been included within, or made the focus of, many risk and vulnerability assessments \cite{ogden2012prism}  \cite{gandini2020holistic} \cite{d2020flood} \cite{miranda2019simplified}. The flood risk of a heritage asset is closely aligned to the proximity of a site to bodies of water, as well as the drainage capacity of local land and wastewater pipes. 

Water ingress from a leaking roof (or rainwater goods), generating staining and humidity problems, is also a common occurrence, though exactly how often these events happen and how much damage is sustained per incident has not been investigated. The substantial number of projects within the MEND and PBIF funding programmes focused upon the replacement of faulty roofs suggests that the deterioration of these elements is a frequent issue experienced within historic buildings, many of which have highly complex roof arrangements comprised of a range of historic materials, making them vulnerable to damage \cite{cassar2003climate}. The incidences of flooding and water ingress are likely to increase in many places around the world as climate change exacerbates the number of intense rainfall events \cite{martel2021climate} \cite{orr2018wind} and storm surges \cite{bevacqua2020more}.

The short list of specific risk scenarios presented in Table \ref{tab:degradation_processes} are only a very limited selection of all the different types of risk that a heritage site has to contend with. To build a fully comprehensive model, a much more expansive collection of risk phrases (and associated metrics), including many coinciding specific risk scenarios for each of the hazards below and other agents of deterioration, e.g., earthquake, pests, terrorism, or breakages, would have to be obtained.
\begin{table}[h]
  \centering
  \caption{Example of adverse events and how they are characterised in the model}
  \label{tab:degradation_processes}
  \begin{tabular}{p{4cm}p{2cm}p{2cm}p{2cm}}
    \toprule
    \textbf{Process Description} & \textbf{Fraction Affected (Range)} & \textbf{Condition Loss (Range)} & \textbf{Mean Time (Range)} \\
    \midrule
    Serious fire where fire service is called in a single room & 0.2 - 0.6 & 100 - 100 & 200 - 600 \\
    Serious incident of heritage crime where high value stolen & 0.002 - 0.006 & 100 - 100 & 5 - 10 \\
    Flood from overflowing drains affecting several ground rooms & 0.02 - 0.06 & 20 - 50 & 1 - 5 \\
    Roof leak in heavy rain causing damage to interiors & 0.02 - 0.06 & 0.6 - 2 & 0.02 - 0.06 \\
    \bottomrule
  \end{tabular}
\end{table}

\subsection{Modelling time to failure with the Weibull distribution}

At every time step of the simulation, an adverse event can occur or not. A way to model this is to sample the time to the next event from a probability distribution. To that end, we adopt the Weibull distribution. This distribution is commonly used in reliability engineering and survival analysis to model the failure rates of mechanical and electronic systems, among other applications. 

The Weibull distribution is characterized by two parameters: \(\lambda\), the scale parameter, and \(k\), the shape parameter. The probability density function (PDF) of the Weibull distribution is given by:

\[ f(x; \lambda, k) = \begin{cases} \frac{k}{\lambda} \left(\frac{x}{\lambda}\right)^{k-1} e^{-(x/\lambda)^k} & x \geq 0 \\ 0 & x < 0 \end{cases} \]

In this equation, \(x\) represents the random variable (e.g., the time to an adverse event), \(\lambda\) represents the scale parameter, which determines the characteristic mean time before an adverse event, and \(k\) represents the shape parameter, which determines the shape of the distribution. 

\begin{itemize}
    \item For \( k < 1 \), the hazard rate decreases over time, indicating that the probability of an event decreases as time progresses.
    \item For \( k = 1 \), the hazard rate remains constant over time.
    \item For \( k > 1 \), the hazard rate increases over time, indicating that the probability of an event increases as time progresses.
\end{itemize}

\begin{figure}[h]
    \centering
    \includegraphics[width=0.8\textwidth]{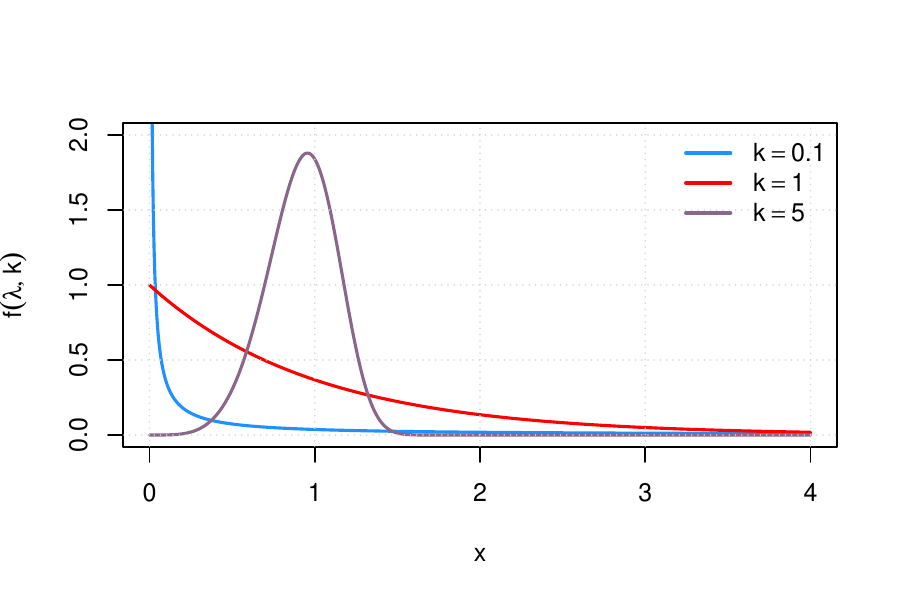}
    \caption{Probability density function of the Weibull distribution for different values of \(k\) and \(\lambda = 1 \).}
    \label{fig:weibull}
\end{figure}

The effect of k can be seen in Figure \ref{fig:weibull}. This implementation of the model uses \( k = 1 \) \Hindex{index:hazard}, even though it is conceivable that in some well known hazards in heritage \( k > 1 \). However, this has never been measured. For example, the older a book is, the more likely a piece will break during handling.

In Figure  \ref{fig:weibull}, when \( k > 1 \), the hazard rate function initially increases with time and then eventually decreases. This characteristic is known as the "bathtub curve" and is commonly observed in reliability engineering.

This curve makes sense in an engineering context, but it remains to be seen if the same logic applies in heritage systems. The idea is that, initially, the hazard rate increases with time because the failure rate accelerates due to wear-out mechanisms. This is often seen in engineering applications in the early life of a product when there is a higher likelihood of defect  due to manufacturing imperfections or stress on components. 

As time progresses, the hazard rate peaks and then declines. This decline occurs because as weaker components fail early in the life of the system, the surviving components tend to be more robust, leading to a decrease in the failure rate. This phase is often referred to as the "random failures" period, where failures occur due to random events rather than wear-out mechanisms.

\subsubsection{Monte Carlo sampling of adverse events}

Within the model, the risk parameters of the ABC framework are used to simulate adverse events and update the condition of affected agents:

\begin{itemize}
    \item When an adverse event occurs, a function is called to select a random subset of agents based on the fraction of the collection affected parameter. This function determines the number of agents affected by the event.
    
    \item For each affected agent, their condition is updated by subtracting a random value within the specified range of the extent of impact on the condition of objects parameter. This simulates the degradation or damage caused by the event to the affected agents.
\end{itemize}

This way of proceeding is usually referred to as Monte Carlo sampling. The properties are assumed to be normally distributed and that the ranges listed in the ABC model correspond to 95$\%$ confidence intervals \Hindex{index:uncertainty}.

\subsection{Agent-Based Model Description}

All the features listed below are put togeher in an  agent-based model that simulates the behavior of a population of agents over time. The main steps of the model are as follows:

\begin{enumerate}
    \item \textbf{Initialization:} The model initializes the population of agents with their initial conditions. Each agent has a condition \( C \) that follows a normal distribution with a mean and standard deviation given by the user, capped between 0 and 100. Information on the current condition of the collection can be introduced in this step.
    
    \item \textbf{Simulation Loop:} The model iterates over multiple years, simulating the behavior of the population for each year.
    
    \item \textbf{Continuous Degradation:} For each year, the condition of each agent is decreased by a degradation rate, calculated from  T and RH conditions. In the first implementation of the model this is calculated with the Strlic damage function, representing continuous degradation over time. It is trivial to add other models, which may apply to different subsets of the population of agents.
    
    \item \textbf{Adverse Events:} The model predicts the occurrence of adverse events using a Weibull distribution. For each year, a random mean time before an adverse event is generated, and a time until event occurrence is sampled from the Weibull distribution. If the sampled time is less than or equal to 1 year, an adverse event is simulated for a fraction of the population, reducing their condition accordingly.
    
    \item \textbf{Percentage of Objects in Good Condition:} After each year, the model calculates the percentage of objects in good condition (condition \( C > 0 \)) and records it.
    
    \item \textbf{Repetition:} The simulation is repeated multiple times to capture the variability of the system. Only 10 runs are usually enough to reveal a characteristing pattern of decay for the collection.
    
    \item \textbf{Analysis:} After all simulations are completed, the model analyzes the results, including histograms of initial and final conditions, time series of the percentage of objects in good condition, and presenting an overall collection lifetime calculated as the average time to reach 1\% of the agents in good condition.
    
\end{enumerate}

\subsection{Implementation in R}

Table \ref{tab:user_inputs_outputs} describes the main inputs and outputs of the first implementation of the model. The code is mostly reproducible following the information given in the previous sections, except for some non-trivial decisions. The truncnorm package generates random numbers from a truncated normal distribution. It is used to initialize the conditions of the agents, ensuring that the starting values are within a realistic and predefined range (i.e. that no agent has a condition above 1 or below 0).

\begin{table}[h]
    \centering
    \begin{tabular}{ p{1.4cm} p{3.2cm} p{6.5cm} }
        \hline
        \textbf{Category} & \textbf{Parameter} & \textbf{Description} \\
        \hline
        \multirow{11}{*}{\textbf{Inputs}} 
         & \texttt{num\_agents} & Number of agents (population size) \\
         & \texttt{num\_years} & Number of years to simulate \\
         & \texttt{num\_simulations} & Number of simulation runs \\
         & \texttt{lower\_bound} & Lower bound of the initial condition distribution \\
         & \texttt{upper\_bound} & Upper bound of the initial condition distribution \\
         & \texttt{mean} & Mean of the initial condition distribution \\
         & \texttt{sd} & Standard deviation of the initial condition distribution \\
         & \texttt{deg\_processes} & List of degradation processes with parameters \\
         & \texttt{T} & Temperature for degradation rate calculation \\
         & \texttt{RH} & Relative humidity for degradation rate calculation \\
         & \texttt{pH} & pH level for degradation rate calculation \\
         & \texttt{DP0} & Initial degradation potential for degradation rate calculation \\
        \hline
        \multirow{5}{*}{\textbf{Outputs}}
         & \texttt{all\_conditions} & Final conditions of agents for each simulation \\
         & \texttt{all\_percentage\_good} & Percentage of agents in good condition over time \\
         & \texttt{time\_to\_1\_percent} & Time taken for the percentage of objects in good condition to drop to 1\% \\
         & \texttt{average\_time} & Average time to reach 1\% good condition across all simulations \\
         & \texttt{sd\_time} & Standard deviation of the time to reach 1\% good condition \\
        \hline
        \multirow{4}{*}{\textbf{Plots}}
         & \texttt{hist} & Histogram of initial and final agent conditions \\
         & \texttt{time\_series\_plot} & Time series plot of percentage of objects in good condition \\
        \hline
    \end{tabular}
    \caption{User Inputs and Main Outputs}
    \label{tab:user_inputs_outputs}
\end{table}

For each year in the simulation, a time until the next event occurrence is sampled from a Weibull distribution with shape parameter \(1\) (indicating an exponential distribution) and scale parameter equal to the mean time between adverse events specified by the users.

The following R code snippet illustrates the use of the `rweibull` function to sample the time until the next event occurrence:

\begin{verbatim}
mean_time <- runif(1, process$mean_time[1], process$mean_time[2])
time_until_event <- rweibull(1, shape = 1, scale = mean_time)
\end{verbatim}

In this snippet, \verb|mean_time| is a random mean time before an adverse event sampled from the specified range. This accounts for the uncertainty in the time to failure. The \verb|mrweibull| function then generates a random deviate from a Weibull distribution with shape parameter \(1\) and scale parameter equal to the sampled mean time. This sampled time until event occurrence determines whether an adverse event will happen in the current year.

\section{Example results}

\begin{figure}[h]
    \centering
    \includegraphics[width=0.7\textwidth]{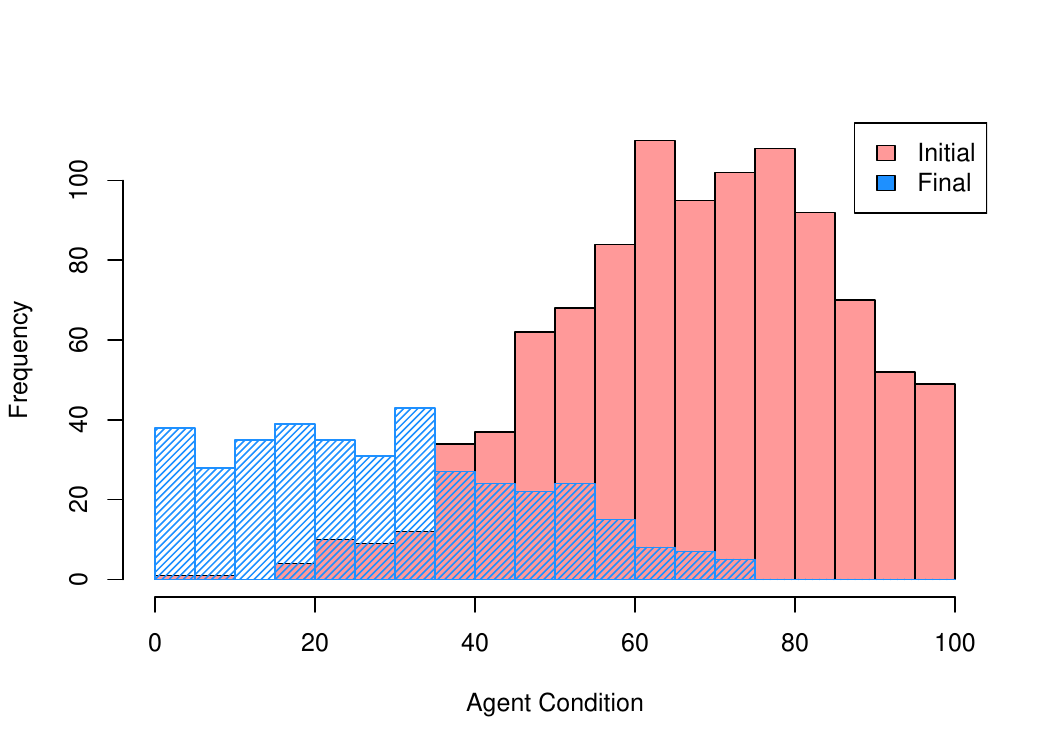}
    \caption{Histogram of condition during a simulation}
    \label{fig:hist}
\end{figure}

The main outcome of the model is the decay in condition of the collection as a consequence of all the combined degradation processes. Figure \ref{fig:hist} shows two histograms of condition, at the start of the process and after 200 years. The mean condition of the collection gradually displaces to the left. When an object reaches a contingent condition of 0, it "falls off" from the simulation (it is deaccessioned, conserved, or otherwise receives some action or investment). 

\begin{figure}[h]
    \centering
    \includegraphics[width=1\textwidth]{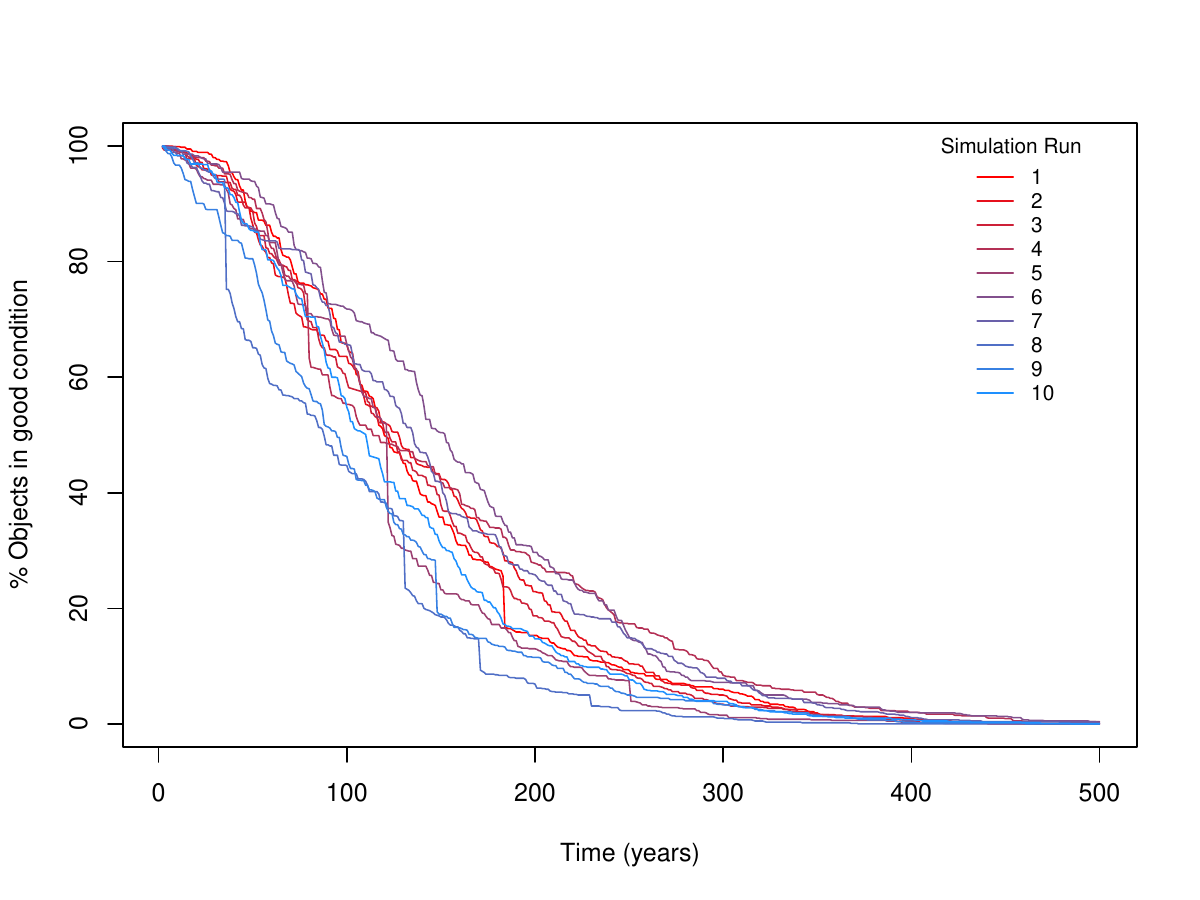}
    \caption{Simulation example showing 10 runs. Some runs exhibit large impact risks while others show many low impact risks, yet overall behaviors are very similar.}
    \label{fig:example}
\end{figure}

At every time step, the model counts how many objects remain in good condition. This allows visualisations like Figure \ref{fig:example}, which shows the evolution of the percentage of objects in good condition over time for 10 example simulation runs. A simulation run represents one possible future lifetime, marked by a series of random accidents. The model should run as many times as necessary to express all the alternative futures for a collection. While nothing prevents us from studying hundreds or thousands of futures, it is interesting to note that in practice, 10 runs are already enough to cover most of the variability in outcomes. Models with more degradation processes may require more runs.

The decay dynamics shown in Figure \ref{fig:example} have some interesting features. The most important observation is that the collection degradation pathways tend to display similar shapes and timelines. This convergence happens even if some futures are more unlucky than others. For example, the simulation run on the left experiences several fires, while the one on the right experiences only minor accidents. And yet, the final outcome, and the overall lifetime, are not as different as could be expected. This occurs because an accumulation of small accidents can be as destructive as a few large accidents. Note that this emergent behaviour is caused only by a list of 4 adverse events (Table \ref{tab:degradation_processes}).

The non-linearity of the decay patterns is due mostly to the initial distribution of conditions. Some non-linearity is also caused by the underlying damage function that continuously erodes the condition. In other implementations of the model, further non-linearity could be introduced by adding self-reinforcing degradation processes (for example, when something is a bit broken it breaks more easily, or when acid degradation starts, it accelerates). Such effects would have an effect on the shape of the distribution of conditions, which would change during the simulation. 

\begin{table}[h]
\centering
\begin{tabular}{l p{2.5cm} p{2cm} p{2cm} }
\hline
\textbf{Condition}                        & \textbf{Collection Lifetime (time to reach 1\%)} & \textbf{Standard Deviation of Lifetime} & \textbf{Max Lifetime} \\ \hline
All degradation processes                 & 377                                             & 21.51                                   & 422                   \\ \hline
Without chemical degradation*             & 450.22                                          & 32.49                                   & 490                   \\ \hline
Without fire*                             & 405.2                                           & 20.2                                    & 434                   \\ \hline
Without theft                             & 369.9                                           & 37.39                                   & 419                   \\ \hline
Without flooding*                         & 792.9                                           & 32.42                                   & 833                   \\ \hline
Without leaking*                          & 417                                             & 26.66                                   & 450                   \\ \hline
\end{tabular}
\caption{Collection Lifetime, Standard Deviation of Lifetime, and Max Lifetime under different conditions. Rows marked with an asterisk (*) show a statistically significant difference compared to "All degradation processes".}
\label{tab:ablation}
\end{table}

One strength of this model is the ability to compare the consequences of degradation processes, regardless of whether they are probabilistic or continuous. One way to do this is what modellers call an "ablation study". In other words, removing one factor at a time in order to observe the effect on the overall outcome. Table \ref{tab:ablation} compares different scenarios where degradation processes have been removed. We can see, for example, that removing chemical degradation increases the lifetime by a bit more than 100 years. Removing flooding has the biggest effect, almost doubling the lifetime. Of course, these results are only as good as the input data. We should not conclude from this that flooding is the most destructive process to collections. Rather, we should use this evidence to critically evaluate our estimate of flooding risk. Have we overestimated its frequency, impact or capacity to reduce value? This is how this method may be helpful to fine-tune the the outcomes of an ABC risk assessment: by allowing us to visualise the consequences of our estimates. 

The model also doubles as an uncertainty propagation analysis. Because risks are defined with ranges, rather than a single value, the model outputs a spectrum of possible outcomes. The standard deviations included in Table \ref{tab:ablation} are produced by averaging the results of all the simulation runs. This helps identify in which cases the effect of a risk is not statistically significant. In this example, removing theft does not result into a statistically significant improvement on the lifetime. 

\subsection{Extensions to the model}

\subsubsection{Incorporate Diverse Degradation Functions}

A number of damage functions are developed enough to be added. The interaction between different degradation mechanisms can be modeled to account for synergistic effects. For example, chemical degradation might make materials more susceptible to physical wear, or high humidity could exacerbate both biological and chemical degradation processes.

\begin{figure}[h]
    \centering
    \includegraphics[width=0.7\textwidth]{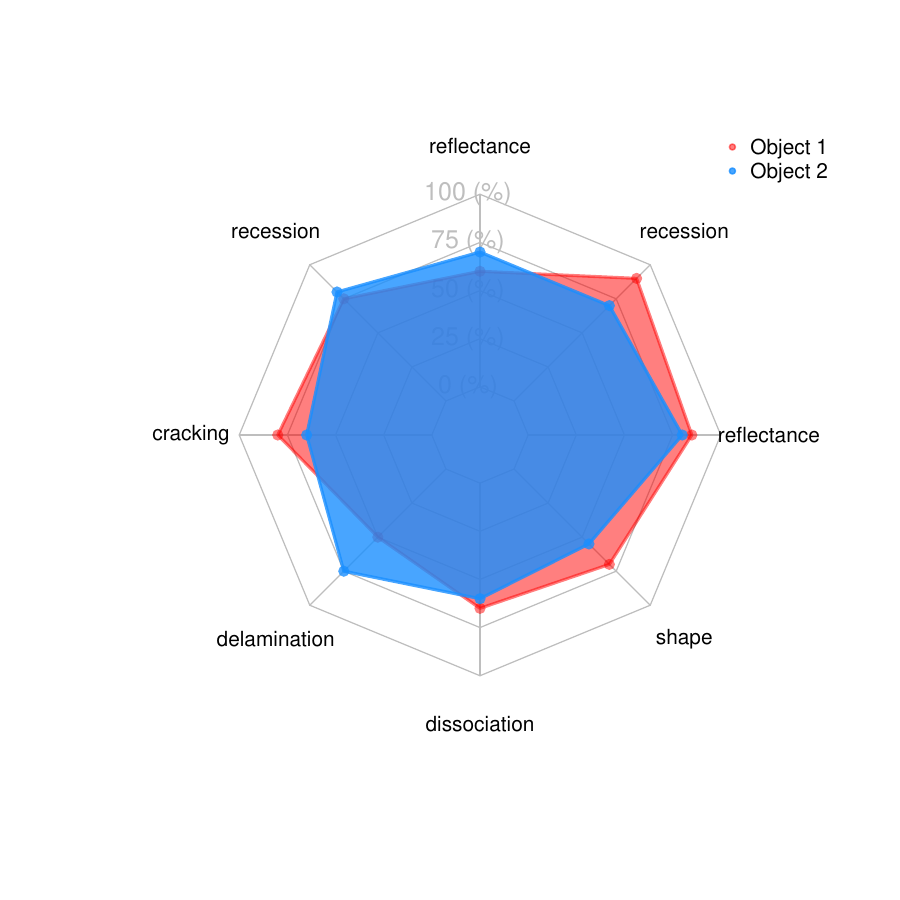}
    \caption{Example of how the model can include different pathways towards condition reduction}
    \label{fig:spider}
\end{figure}

\subsubsection{Dynamic Environmental Conditions}

Rather than relying on static values for parameters like temperature, relative humidity (RH), the model can incorporate variable environmental data. Seasonal variations can be introduced to simulate the cyclical nature of environmental conditions, which is especially interesting if mechanical damage is added. 

\subsubsection{Agent-Based Model Enhancements}

The first step to enhance the agent-based model involves increasing the heterogeneity among agents. In particular, each agent can degrade in a different way. Even further, it is possible to define several properties for each agent. Figure \ref{fig:spider} shows an invented example, inspired by the spider plots used for sensory profiles, i.e., in wine tasting. In this case, each separate agent has 8 dimensions of condition. All of them can be lost, and the agent would become "unacceptably degraded" when one of them reaches the threshold. The second enhancement to the model would be to add interactions between agents. Early experimentation with this idea \cite{duran2024end} has involved the interaction between book-agents and visitor-agents, which causes accelerated degradation due to handling. A third enhancement would be to consider sub-spaces within collections, such as different rooms or storage areas. 

\subsubsection{Scenario Planning }

Scenario planning would extend the model's applicability by exploring potential future scenarios and their impacts on the collection. This involves running simulations under different assumptions, such as varying climate conditions, funding levels for conservation, or changes in storage environments. Initial research has involved evaluating the cost of delaying the decision to deacidify a collection \cite{duran2021comparison}. 

\section{Implications for future research}

Can the principles of Collections Demography be extended to cover any heritage typology? The answer depends on the testing of six hypotheses, which sustain the model presented in this paper:

\begin{itemize}
    \item H\ref{index:exist} Threshold Existence: There exist identifiable thresholds for unacceptable change in a wide diversity of heritage typologies. This could be difficult to define for heritage typologies that have complex or diverse social uses. 
    \item H\ref{index:first} Sharp Response: The perception and response to condition change in heritage objects increase sharply beyond a certain threshold. If this hypothesis does not hold, the definition of lifetime will be more arbitrary. 
    \item H\ref{index:proportional} Proportional Degradation: An absolute condition can be defined in a way that is directly proportional to one or more key measurable properties of heritage objects (e.g., color, strength). This could be complicated in degradation processes which lead to multi-dimensional phenomena that is not characterised with a single metric, such as crack networks.
    \item H\ref{index:ABC} Risk Forecasting: The ABC risk assessment model or similar frameworks can evolve into a quantitative tool for forecasting risks associated with heritage objects. This is achievable with a combination of data and expertise.
    \item H\ref{index:hazard} Time Distribution: A Weibull distribution effectively models hazard rates or failure rates of heritage objects. Exploring this requires data.  
    \item H\ref{index:uncertainty} Modeling Uncertainty: The uncertainty in the severity of adverse events is measurable and can be modelled, for example with a normal distribution. Exploring this requires data. 
\end{itemize}

The model presented can be useful even before these six hypothesis are thoroughly investigated (or indeed even if one or two are disproved). The uses of the model are:

\begin{itemize}
    \item To compare the consequences of probabilistic and continuous degradation.
    \item To assess critically the quality of risk assessment estimations, by checking if the long-term impact of estimated risks is realistic in comparison with other processes. 
    \item To study the propagation of different types of uncertainty to the final lifetime estimation. For example, comparing the uncertainty caused by measurement errors (e.g. $\pm$ 3$\%$ RH) with the uncertainty caused by expert estimations of unknown parameters.
\end{itemize}

While the six hypotheses are necessary for a \emph{universal} model, there is a high potential to use this approach in specific contexts. We know this works for paper collections. What other collection types could benefit from this type of analysis with our current level of knowledge?

\section*{Model Availability}

The R implementation of the collection degradation model described in this document is available on GitHub:

\begin{center}
\href{https://github.com/jgraubove/generaldegradation}{\textbf{https://github.com/jgraubove/generaldegradation}}
\end{center}

Feel free to clone the repository, explore the code, and use it for your own simulations. Contributions and feedback are welcome!

\section*{Funding}

This research has been funded by DCMS and the AHRC grant AH/Y000439/1.

\bibliographystyle{naturemag}
\bibliography{references}

\end{document}